\documentclass[12pt]{article}
\usepackage[dvips]{color}
\usepackage{amsmath}
\usepackage{graphicx}
\usepackage{subfigure}
\usepackage{epsfig}
\usepackage{amsmath}
\usepackage{cite}
\usepackage{color}
\usepackage{subfigure}

\begin{document}
\begin{center}
\Large{\bf Bardeen Black Hole Thermodynamics\\ from Topological Perspective}\\
 \small \vspace{1cm}
 {\bf Jafar Sadeghi $^{\star}$\footnote {Email:~~~pouriya@ipm.ir}}, \quad
 \small \vspace{0.3cm}{\bf Saeed Noori Gashti$^{\dag,\star}$\footnote {Email:~~~saeed.noorigashti@stu.umz.ac.ir}}, \quad
 {\bf Mohammad Reza Alipour $^{\star}$\footnote {Email:~~~mr.alipour@stu.umz.ac.ir}}\quad\\
 {\bf Mohammad Ali S. Afshar $^{\star}$\footnote {Email:~~~m.a.s.afshar@gmail.com}}\quad
\\
\vspace{0.5cm}$^{\star}${Department of Physics, Faculty of Basic
Sciences,\\
University of Mazandaran
P. O. Box 47416-95447, Babolsar, Iran}\\
\vspace{0.5cm}$^{\dag}${School of Physics, Damghan University, P. O. Box 3671641167, Damghan, Iran}
\small \vspace{1cm}
\end{center}
\begin{abstract}
In this paper, we use the generalized off-shell Helmholtz free energy method to explore the thermodynamic properties of Bardeen black holes (BD BHs) from a topological perspective based on Duan's topological current $\phi$-mapping. We consider various structures of BD BHs, including regular BD-AdS BHs, BD-AdS BHs in Kiselev's model of quintessence, BD BHs in massive gravity (MG), and BD BHs in 4D Einstein-Gauss-Bonnet (EGB) gravity. We demonstrate that these BHs have one topological classification (TC), i.e., TC is +1 for all BHs considered, and the addition of MG or GB terms, etc., does not change the topological numbers.\\\\
Keywords: Topological Classification, Bardeen Black Holes, Helmholtz Free Energy
\end{abstract}
\newpage
\tableofcontents
\section{Introduction}
To gain insight into quantum gravity, researchers consider various approaches, including the study of BHs and their thermodynamic characteristics, which is a promising tool \cite{1,2}.
This idea involves combining classical thermodynamics and statistical mechanics to interpret the entropy of BHs as the number of quantum
degrees of freedom near the event horizon. The thermodynamics of BHs have been investigated in different frameworks, and the results can be found in \cite{1001,1002,1003,1004,1005,1006,1007,1008,1009,1010,1011,1012,1013,1014}.
One important concept in thermodynamics is the Hawking-Page phase transition \cite{3}, which indicates a specific relationship between gravity and thermodynamics.
This idea suggests that there is a phase transition between thermal radiation in the AdS space and BHs, known as confinement/deconfinement phase transitions in
the context of the AdS/CFT correspondence.\\\\
Also, in the thermodynamics of BHs, the study of phase transitions and critical points always attracts the most attention. In this regard, the observed structural similarities between the liquid-to-gas phase transition in van der Waals fluid and the first-order phase transition between small and large BHs are subjects that cannot be easily ignored.\\
Recently, new approaches have been proposed to study and calculate critical points and phase transitions in the thermodynamics of BHs. One of these approaches is the topological approach, which has been mentioned in \cite{7aa,e,f,13}.
To obtain a topological perspective in thermodynamics, one of the best options is to use Duan’s topological current $\phi$ mapping theory.
Wei et al. presented two different approaches to study topological thermodynamics based on temperature and generalized free energy function.\\\\
\textit{The first method} involves studying the temperature function by removing pressure and using the auxiliary and topological parameter $1/\sin \theta$, and then constructing the potential based on these assumptions. For further study, you can see \cite{7aa,e,f,13}.\\\\
\textit{In the second approach}, with the assumption that BHs can be considered as defects in the thermodynamic parameter space, their solutions can be investigated using the generalized off-shell free energy.
In that case, the stability and instability of the obtained BH solutions are determined by positive and negative winding numbers, respectively. Also in this theory, the properties of a field configuration can be determined from the zero points of the field in space, represented by $\phi(\vec{x})|_{\vec{x}=\vec{z}} = 0$ \cite{4,5,6,7}.\\
Wei, Liu, and B. Mann claimed that for different branches of a BH at an arbitrary temperature, the topological charge number, which is the sum of the winding numbers, is a universal number independent of the BH parameters.
In that case, in the range of small and large BHs, the topological charge number depends only on the thermodynamic asymptotic behavior of the BH temperature. So, BHs can be divided into three different categories through the number of topological charges\cite{5}.\\\\
After the introduction of this second method, different BHs were also
studied to investigate topological charges,
which can be briefly mentioned as examples:\\
C. H. Liu and J. Wang showed that the topological number remains constant
and the charged GB BHs in the AdS
space are placed in the same category as the RN-AdS, with the same topological
numbers.
But the obtained results showed that charge dependence was not the only
effective factor, and if we remove the charges, it can be seen that the topological
the number can show dimensional dependence.
This means that the conclusion that a topological number is necessarily a universal number independent of BH parameters is questioned\cite{8}.\\
In another work, Y. Du and Xi. Zhang chose the rotating charged BTZ BH model to
study and after calculation, they found that there are only two topological classes
for BTZ space-time.
They stated that in the special case $Q = J = 0$ in this BH, it has only one zero
point and therefore the total topological number is 1.
Whereas for the spin or charged cases, there must always be two zero points and the
total topological number is zero\cite{14}. Using the same method, Wu investigated the structure of charged Lorentzian Taub-NUT space-times and neutral Lorentzian NUT-charged space-times in 4 dimensions in two separate articles. The results showed that the presence of the NUT parameter did not affect the topological number in the neutral state space-time. Thus, Taub-NUT, Taub-NUT-AdS, and Kerr-NUT can be considered as generic black holes from the perspective of the thermodynamic topological approach\cite{9}. In the second work, Wu found that the presence of the NUT charge as a parameter did not affect the structures in asymptotically flat space-time but, had an effect on asymptotically local AdS space-time \cite{131415}.\\\\
These concepts motivated us to choose the structure of different Bardeen BHs based on this method and present the results of our studies as follows.\\
In section 2, we briefly describe the thermodynamics of BHs from the point of
view of topology using the generalized off-shell Helmholtz free energy method. In
section 3, we first explain BD-BHs and then introduce the different structures of
BD-BHs in 4 sub-sections, namely BD-AdS BHs, regular BD-AdS BHs in the
Kiselev’s model of quintessence (The Kiselev BH and a BH
surrounded by quintessence are two different types of BH solutions. The
Kiselev BH is a static, spherically symmetric solution to Einstein’s
equations with a non-zero cosmological constant and an anisotropic stress-energy
tensor. The anisotropy of the stress-energy tensor leads to a non-zero relative
pressure anisotropy, which distinguishes it from a perfect fluid. On the other hand,
a BH surrounded by quintessence is a solution that includes a scalar field,
which is responsible for the quintessence. The scalar field has an energy density
and pressure that contributes to the stress-energy tensor, and the resulting solution is
not necessarily spherically symmetric. In summary, the Kiselev BH is a
solution with an anisotropic stress-energy tensor, while a BH surrounded by
quintessence is a solution that includes a scalar field \cite{14a}), BD-BHs in MG, and
BD-AdS BHs in 4D EGB Gravity. We then calculate the ZPs and search for the
topological charges of each BH. Finally, we describe the results of our work in
detail in section 4.
\section{Topology of black holes thermodynamics}
To introduce the thermodynamic properties of BHs, various quantities are used. For instance, two different variables, such as mass and temperature, can be used to describe the generalized free energy.
Considering the relationship between mass and energy in BHs, we can rewrite our generalized free energy function as a standard thermodynamic function in the following form \cite{9,11,12},
\begin{equation}\label{1}
\mathcal{F}=M-\frac{S}{\tau}.
\end{equation}
In the above equation, $\tau$ is the Euclidean time period and T (inverse of $\tau$) stands the temperature of the ensemble. The generalized free energy is on-shell only when $\tau= \tau_{H} =\frac{1}{T_{H}}$. According to \cite{11,12}, a vector $\phi$ is constructed as follows,
\begin{equation}\label{2}
\phi=\big(\frac{\partial\mathcal{F}}{\partial r_{H}}-\cot\Theta\csc\Theta\big).
\end{equation}
where $\phi^{\Theta}$ is divergent, the direction of the vector points outward when $\Theta= 0, \pi$. For $r_{H}$ and $\Theta$, we have $0\leq r_{H}\leq\infty$ and $0\leq\Theta\leq\pi$, respectively. A topological current can be defined using Duan's $\phi$-mapping topological current theory as follows \cite{11,12},
\begin{equation}\label{3}
j^{\mu}=\frac{1}{2\pi}\varepsilon^{\mu\nu\rho}\varepsilon_{ab}\partial_{\nu}n^{a}\partial_{\rho}n^{b},\hspace{1cm}\mu,\nu,\rho=0,1,2
\end{equation}
where $n=(n^1, n^2)$, we have $n^1=\frac{\phi^r}{\|\phi\|}$ and $n^2=\frac{\phi^\Theta}{\|\phi\|}$. By Noether's theorem, we know that the resulting topological currents are conserved,
\begin{equation}\label{4}
\partial_{\mu}j^{\mu}=0,
\end{equation}
To calculate the topological number, we re-express the topological current as \cite{9,11,12},
\begin{equation}\label{5}
j^{\mu}=\delta^{2}(\phi) J^{\mu}(\frac{\phi}{x}),
\end{equation}
Where the Jacobi tensor is defined as follows,
\begin{equation}\label{6}
\varepsilon^{ab}J^{\mu}(\frac{\phi}{x})=\varepsilon^{\mu\nu\rho}\partial_{\nu}\phi^{a}\partial_{\rho}\phi^{b}
\end{equation}
The Jacobi vector becomes the usual Jacobi when $\mu=0$, as shown by $J^{0}\big(\frac{\phi}{x}\big)=\frac{\partial(\phi^1,\phi^2)}{\partial(x^1,x^2)}$. Equation (4) indicates that $j^{\mu}$ is only non-zero when $\phi=0$. With some calculations, we can determine the topological number or total charge $W$ in the following form:
\begin{equation}\label{7}
W=\int_{\Sigma}j^{0}d^2 x=\Sigma_{i=1}^{n}\beta_{i}\eta_{i}=\Sigma_{i=1}^{n}\omega_{i}.
\end{equation}
where $\beta_i$ is the positive Hopf index, counting the loops of vector $\phi^a$ in the $\ phi$ space when $x^\mu$ is near the zero point $z_i$. Meanwhile, $\eta_i=\text{sign}(j^0(\phi/x)_{z_i})=\pm 1$. The quantity $\omega_i$ represents the winding number for the $i$-th zero point of $\phi$ in $\Sigma$. Note that the winding number is independent of the region's shape where the calculation occurs. The winding number's value is directly associated with BH stability, where a positive (negative) winding number corresponds to a stable (unstable) BH state.
\section{Topological classification of Bardeen black holes}
BHs are complex and mysterious astronomical objects that are characterized by singularities concealed by event horizons. In the context of general relativity, a singularity is a region where all known physical laws break down, and the gravitational pull diverges in the affected area of spacetime.
To address these unresolved regions, a set of solutions known as "regular BHs" has been instrumental in overcoming this challenge, as they lack singularities even at the origin. Bardeen was the first to obtain a regular, spherically symmetric BH solution, known as the BD BH \cite{15}.
As this solution was not a vacuum solution, a specialized form of the energy-momentum tensor was introduced to achieve a model that satisfied the weak energy condition. Since then, the properties of regular BH solutions have been extensively studied in the context of general relativity.
The BD-BH space-time satisfies the weak energy condition but not the strong energy condition \cite{22,23,24,25,26}. In this study, we will provide a brief overview of four different BD- BH structures before exploring the thermodynamic properties of BHs. These include BD-AdS BHs, regular BD-AdS BHs in Kiselev's model of quintessence, BD-BHs in MG, and BD-AdS BHs in 4D EGB Gravity, from a topological perspective. We will use Duan's topological current $\phi$-mapping theory and the generalized off-shell Helmholtz free energy method. As the existence of ZPs indicates a topological charge, we will determine the topological charge of each BH by identifying the topological charges that align with the winding numbers. Finally, we will compare our results with other works in the literature.
\subsection{Regular Bardeen AdS black holes}
From a topological perspective, we examine the thermodynamics of BD-AdS BHs and explore their topological properties in an extended phase space. The four-dimensional spherically symmetric BD-AdS BH can be described by \cite{27},
\begin{equation}\label{8}
ds^{2}=-f(r)dt^{2}+\frac{dr^{2}}{f(r)}+r^{2}d\Omega^{2}.
\end{equation}
and
\begin{equation}\label{9}
f(r)=1-\frac{2Mr^2}{(q^2+r^2)^{\frac{3}{2}}}+\frac{r^2}{\ell^2}
\end{equation}
where $d\Omega^2$ represents the line element of a unit two-sphere. The mass, entropy, and Hawking temperature of this BH can be calculated,
\begin{equation}\label{10}
M =\frac{\left(1+\frac{8 r^{2} \pi  P}{3}\right) \left(q^{2}+r^{2}\right)^{\frac{3}{2}}}{2 r^{2}},
\end{equation}
\begin{equation}\label{11}
S =\pi r^{2},
\end{equation}
\begin{equation}\label{12}
T=\frac{-2q^2+r_{H}^2+3r_{H}^{4}/\ell^2}{4\pi r_{H}(q^{2}+r_{H}^2)},
\end{equation}
Here, according to equations (1), (9), and (11), we can obtain the generalized Helmholtz free energy for the mentioned BH.
\begin{equation}\label{13}
\mathcal{F}=\frac{\left(1+\frac{8 r^{2} \pi  P}{3}\right) \left(q^{2}+r^{2}\right)^{\frac{3}{2}}}{2 r^{2}}-\frac{\pi  r^{2}}{\tau}
\end{equation}
Based on what was stated in the previous section, the form of the function $\phi(r)$ is determined as follows,
\begin{equation}\label{14}
\phi_{r}=\frac{8 \pi  P \left(q^{2}+r^{2}\right)^{\frac{3}{2}}}{3 r}+\frac{3 \left(1+\frac{8 r^{2} \pi  P}{3}\right) \sqrt{q^{2}+r^{2}}}{2 r}-\frac{\left(1+\frac{8 r^{2} \pi  P}{3}\right) \left(q^{2}+r^{2}\right)^{\frac{3}{2}}}{r^{3}}-\frac{2 \pi  r}{\tau},
\end{equation}
Now, by solving the above equation with respect to $\phi_{r}=\frac{\partial \mathcal{F}}{\partial r}$, we can calculate,
\begin{equation}\label{15}
\tau =\frac{4 \pi  r^{4}}{\sqrt{q^{2}+r^{2}}\, \left(8 \pi  P \,r^{4}-2 q^{2}+r^{2}\right)}
\end{equation}

\begin{figure}[h!]
 \begin{center}
 \subfigure[]{
 \includegraphics[height=5.5cm,width=6cm]{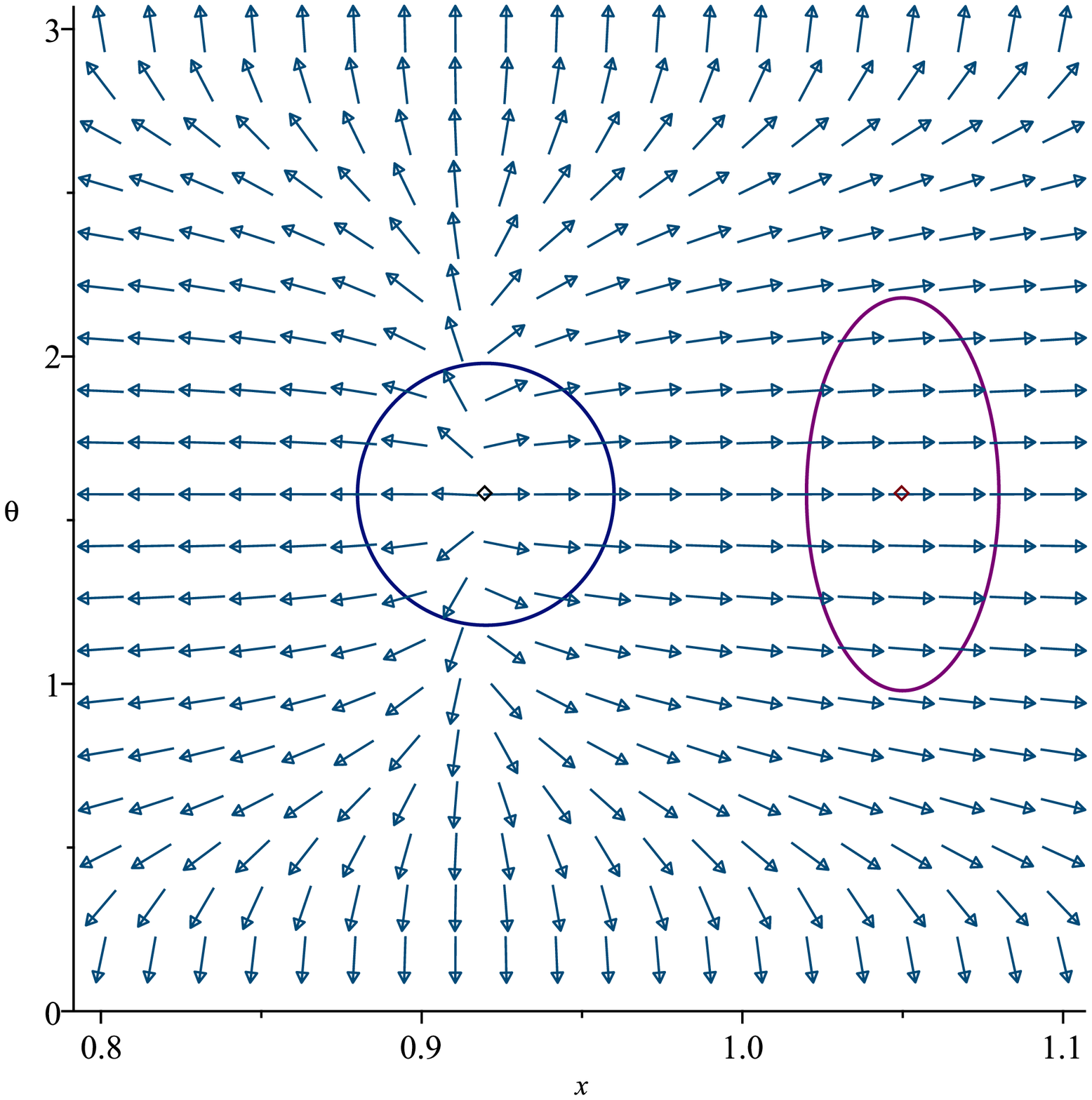}
 \label{1a}}
 \subfigure[]{
 \includegraphics[height=5.5cm,width=6cm]{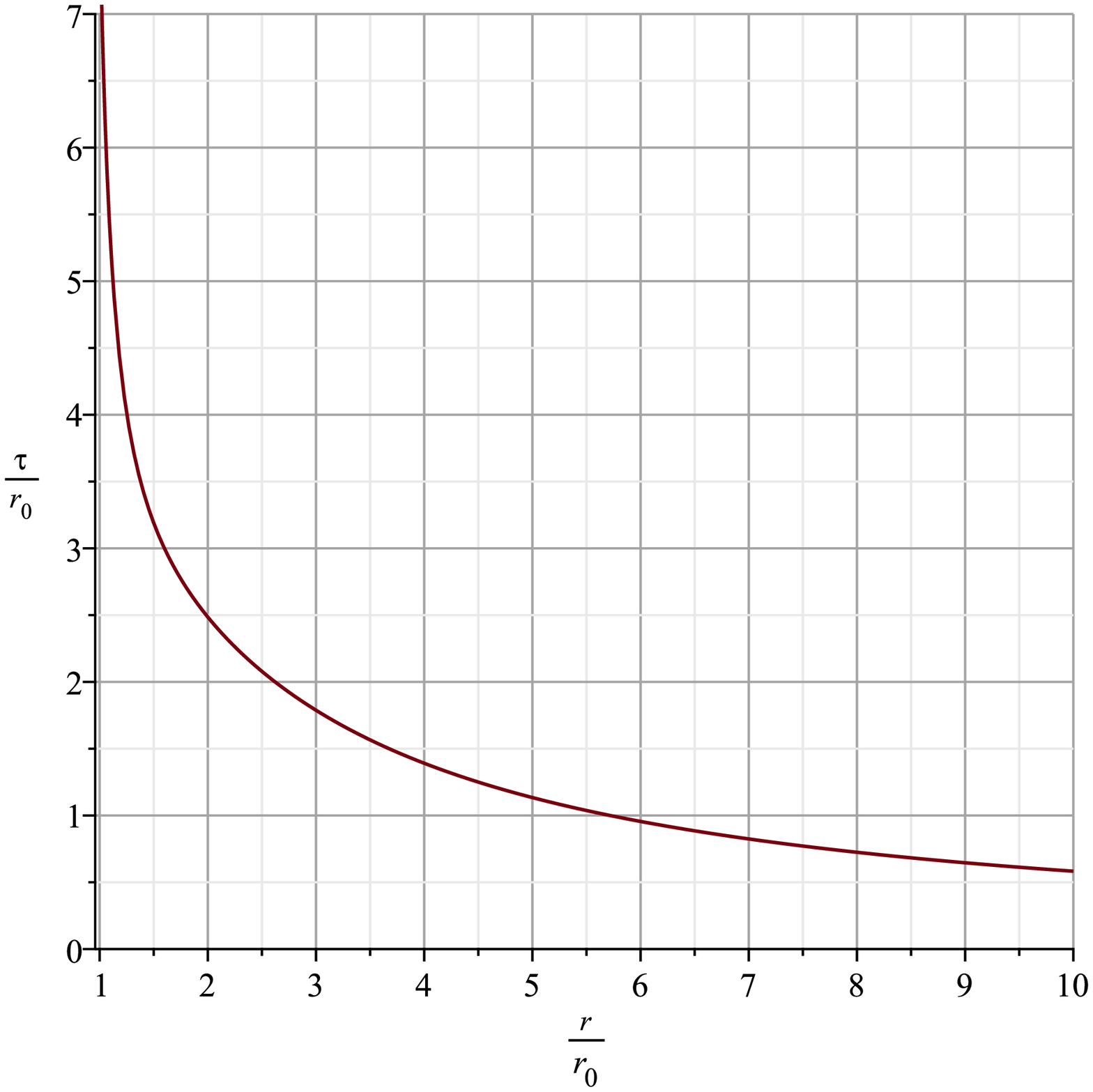}
 \label{1b}}
  \caption{\small{In Fig (1a), the blue arrows represent the vector field $n$ on a portion of the $(x-\theta)$ plane for the regular BD-AdS BH with $(q=r_{0}, P=0.085/r_{0}^{2})$, where $(x=r/r_{0})$. The ZP is located at $(x,\theta)=(0.913, \pi/2)$ on the circle loop ($C_{1}$). The contours $C_1$ (blue loop) and $C_2$ (purple loop) are two closed loops, with $C_1$ enclosing the ZP while $C_2$ does not. The plot of the curve of equation (15) is shown in Fig (1b).}}
 \label{1}
 \end{center}
 \end{figure}
\subsection{Bardeen AdS black holes in Kiselev's model of quintessence}
In this subsection, we discuss the importance of studying the impact of Kiselev's model of quintessence on the BD-BHs. We utilize Kiselev's phenomenological model to create a regular BD BH, which is a type of BH that lacks a singularity but still has a horizon. The regular BD BH is then surrounded by quintessence. Kiselev's model is not a solution of field equations derived from an action coupled to a quintessence field (such as a scalar field). Also, the stress energy-momentum tensor used in Kiselev's paper is anisotropic and hence does not represent perfect fluid \cite{14a}. To obtain the metric that accounts for the effect of quintessence on the regular BD BH, Einstein's equations are solved. This metric can provide insights into how quintessence affects the structure and behavior of BHs. So, we will have \cite{28,29},
\begin{equation}\label{16}
ds^{2}=-f(r)dt^{2}+\frac{dr^{2}}{f(r)}+r^{2}d\theta^{2}+r^{2}\sin^{2}\theta d\phi^{2},
\end{equation}
where
\begin{equation}\label{17}
f(r)=\bigg(1-\frac{2Mr^{2}}{(q^{2}+r^2)^{\frac{3}{2}}}+\frac{r^2}{\ell^2}-\frac{a}{r^{3\omega_{q}+1}}\bigg),
\end{equation}
where $\omega_{q}$ denotes the state parameter. Also, the mass of the BH can be determined by the condition $f(r_+)=0$ at the event horizon $r_+$,
\begin{equation}\label{18}
M =\frac{\left(q^{2}+r^{2}\right)^{\frac{3}{2}} \left(\frac{8 \left(\frac{9}{8 \pi  P}+3 r^{2}\right) \pi  P}{3 r^{2}}-\frac{3 a}{r^{3 \omega_q +3}}\right)}{6}
\end{equation}
The entropy of the BH is given by,
\begin{equation}\label{19}
S =\pi  r^{2}
\end{equation}
Like the calculations in the previous section, here we also calculate the generalized Helmholtz free energy for regular BD-AdS BHs with quintessence, using equations (1), (17), and (19). Therefore, we obtain,
\begin{equation}\label{20}
\mathcal{F}=\frac{\left(q^{2}+r^{2}\right)^{\frac{3}{2}} \left(\frac{8 \left(\frac{9}{8 \pi  P}+3 r^{2}\right) \pi  P}{3 r^{2}}-\frac{3 a}{r^{3 \omega_q +3}}\right)}{6}-\frac{\pi  r^{2}}{\tau}
\end{equation}
So, with respect to the above equation, we can calculate,
\begin{equation}\label{21}
\tau =\frac{4 \pi  r^{4} r^{3 \omega_q +3}}{\sqrt{q^{2}+r^{2}}\, \left(8 \pi  r^{3 \omega_q +3} P \,r^{4}+3 a q^{2} \omega_q   r^{2}+3 a \omega_q  r^{4}+3 a \,q^{2} r^{2}-2 r^{3 \omega_q +3} q^{2}+r^{3 \omega_q +3} r^{2}\right)}
\end{equation}

\begin{figure}[h!]
 \begin{center}
 \subfigure[]{
 \includegraphics[height=5.5cm,width=6cm]{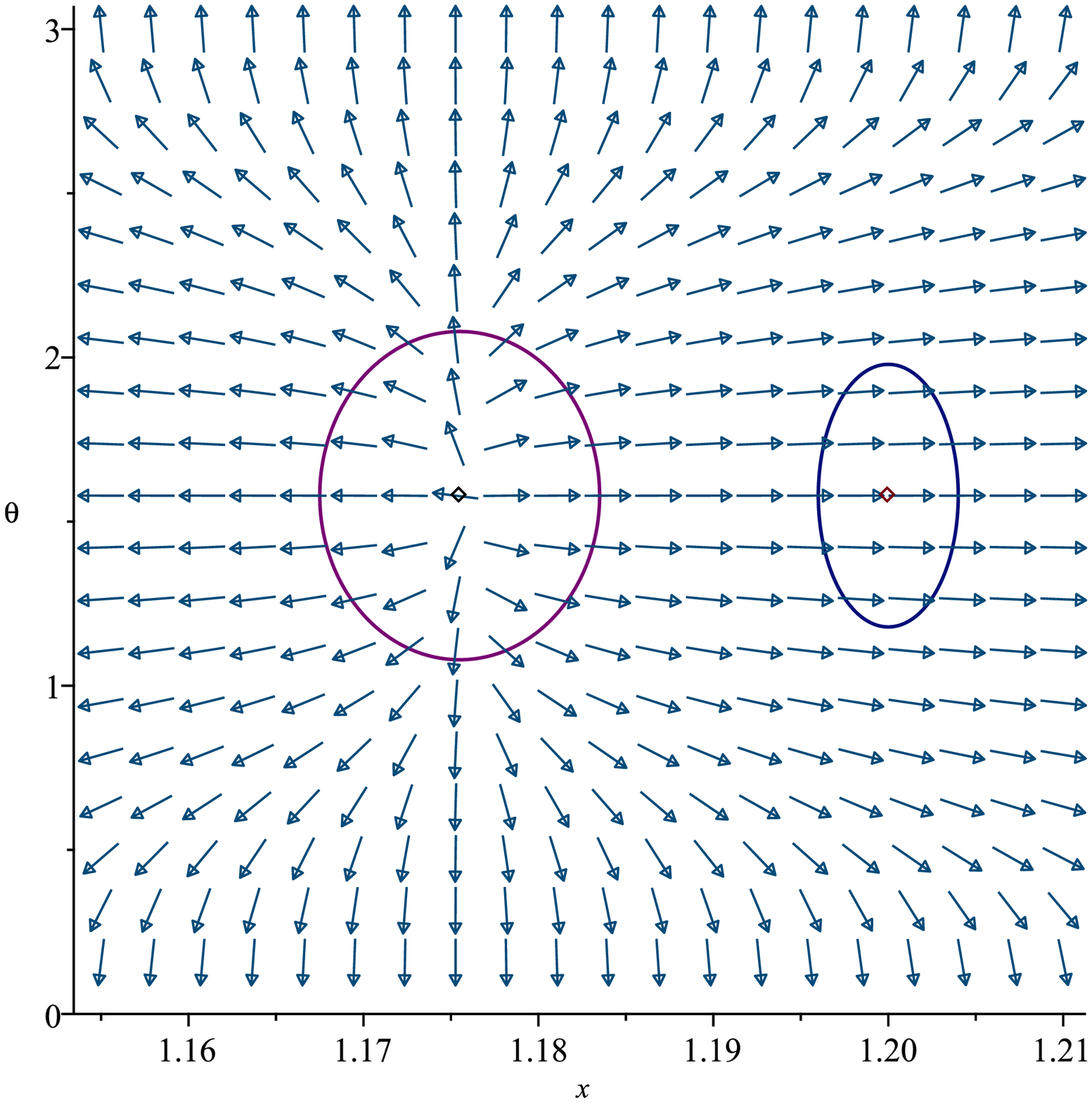}
 \label{2a}}
 \subfigure[]{
 \includegraphics[height=5.5cm,width=6cm]{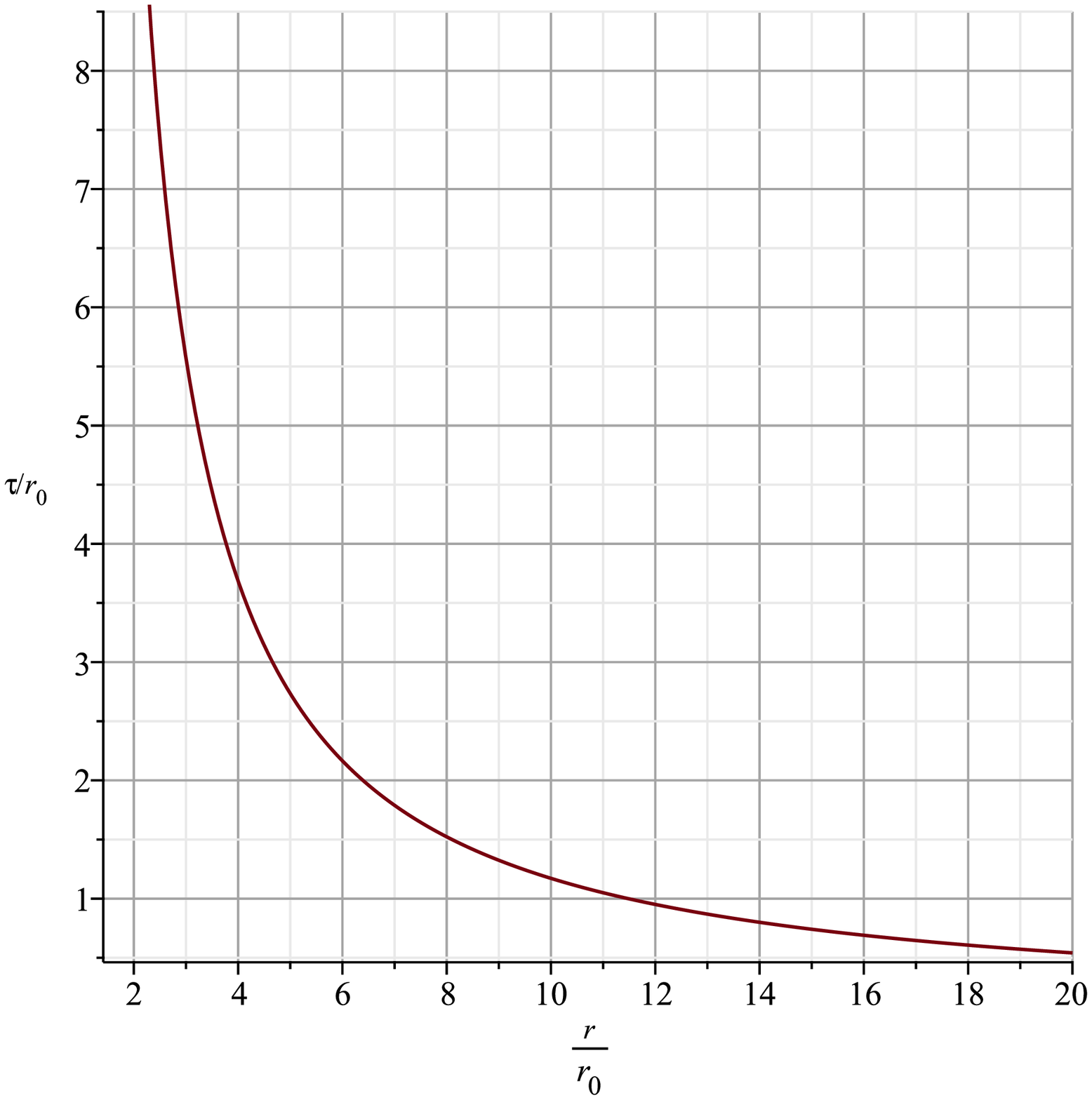}
 \label{2b}}
  \caption{\small{ In Fig (2a), the blue arrows represent the vector field $n$ on a portion of the $(x-\theta)$ plane for regular BD-AdS BHs with quintessence, with respect to $q=r_{0}, \omega_q=-2/3, a=1/r_{0}, P=0.05/r_{0}^{2}$ where $(x=r/r_{0})$. The ZP is located at $(x,\theta)=(1.175, \pi/2)$ in the circle loop ($C_{1}$). The contours (purple loop) and $C_2$ (blue loop) are two closed loops, with $C_1$ enclosing the ZP while $C_2$ does not. The plot of the curve of equation (21) is shown in Fig (2b).}}
 \label{2}
 \end{center}
 \end{figure}
 In Fig (2a), the blue arrows represent the vector field $n$ on a portion of the $(x-\theta)$ plane for regular BD-AdS BHs with quintessence, with respect to $q=r_{0}, \omega_q=-2/3, a=1/r_{0}, P=0.05/r_{0}^{2}$ where $(x=r/r_{0})$. The ZP is located at $(x,\theta)=(1.175, \pi/2)$ in the circle loop ($C_{1}$). The contours (purple loop) and $C_2$ (blue loop) are two closed loops, with $C_1$ enclosing the ZP while $C_2$ does not. The plot of the curve of equation (21) is shown in Fig (2b), where each point on the curve corresponds to an on-shell BH.

\subsection{Bardeen AdS black holes in massive gravity}
Continuing from the previous process, in this subsection, we will explore the four-dimensional BD-BH in the presence of MG, which is characterized by parameters $m$ and coefficients $c$. Our main objective is to investigate the impact of the new coefficients on the number and type of topological charges and, consequently, the thermodynamic changes in the system. We have \cite{30},
\begin{equation}\label{22}
f(r)=1-\frac{2Mr^2}{(r^2+q^2)^{\frac{3}{2}}}+\frac{r^2}{\ell^2}+m^{2}\big(c^2c_2+\frac{cc_1 r}{2}\big),
\end{equation}
The solution depicts the 4D AdS-BD BH with MG, characterized by the mass $M$, cosmological constant $\Lambda=-3/\ell^2$, magnetic charge $q$, the mass parameter $m$, and constants $c_i$. The thermodynamic quantities for the BD-AdS BH with MG, including entropy ($S$) and mass, can be calculated,
\begin{equation}\label{23}
M =\frac{\left(q^{2}+r^{2}\right)^{\frac{3}{2}} \left(1+\frac{8 r^{2} \pi  P}{3}+m^{2} \left(c^{2} \beta +\frac{1}{2} c \alpha  r \right)\right)}{2 r^{2}}
\end{equation}
\begin{equation}\label{24}
S =\pi  \left(\left(-2 q^{2}+r^{2}\right) \sqrt{q^{2}+r^{2}}+3 q^{2} r \ln \! \left(r +\sqrt{q^{2}+r^{2}}\right)\right)
\end{equation}
Furthermore, we compute the generalized Helmholtz free energy $\mathcal{F}$ as follows,
\begin{equation}\label{25}
\begin{split}
&\mathcal{F} =\frac{\left(q^{2}+r^{2}\right)^{\frac{3}{2}} \left(1+\frac{8 r^{2} \pi  P}{3}+m^{2} \left(c^{2} \beta +\frac{1}{2} c \alpha  r \right)\right)}{2 r^{2}}\\
&-\frac{\pi  \left(\left(-2 q^{2}+r^{2}\right) \sqrt{q^{2}+r^{2}}+3 q^{2} r \ln \! \left(r +\sqrt{q^{2}+r^{2}}\right)\right)}{\tau}
\end{split}
\end{equation}
We obtain the $\tau$ with respect to equation (25). So, we will have,
\begin{equation}\label{26}
\begin{split}
&\mathcal{A}=12 \pi  r^{3} \left(\sqrt{q^{2}+r^{2}}\, \ln \! \left(r +\sqrt{q^{2}+r^{2}}\right) q^{2}+q^{2} r +r^{3}\right)\\
&\mathcal{B}=-\alpha  c \,m^{2} q^{4} r +\alpha  c \,m^{2} q^{2} r^{3}+2 \alpha  c \,m^{2} r^{5}-4 \beta  c^{2} m^{2} q^{4}-2 \beta  c^{2} m^{2} q^{2} r^{2}+2 \beta  c^{2} m^{2} r^{4}\\
&+16 \pi  P \,q^{2} r^{4}+16 \pi  P \,r^{6}-4 q^{4}-2 q^{2} r^{2}+2 r^{4}\\
&\tau =\frac{\mathcal{A}}{\mathcal{B}}
\end{split}
\end{equation}
\begin{figure}[h!]
 \begin{center}
 \subfigure[]{
 \includegraphics[height=5.5cm,width=6cm]{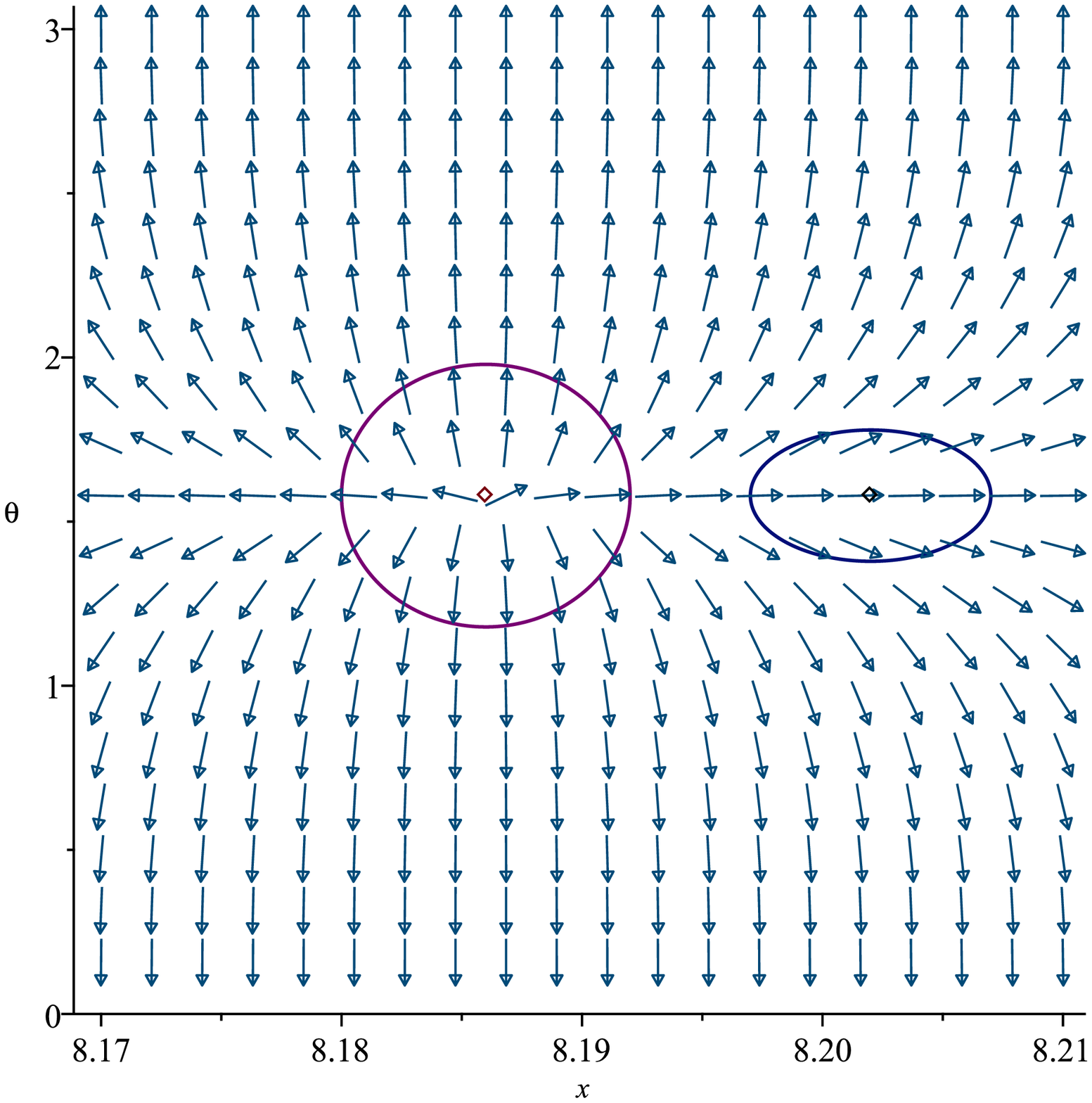}
 \label{3a}}
 \subfigure[]{
 \includegraphics[height=5.5cm,width=6cm]{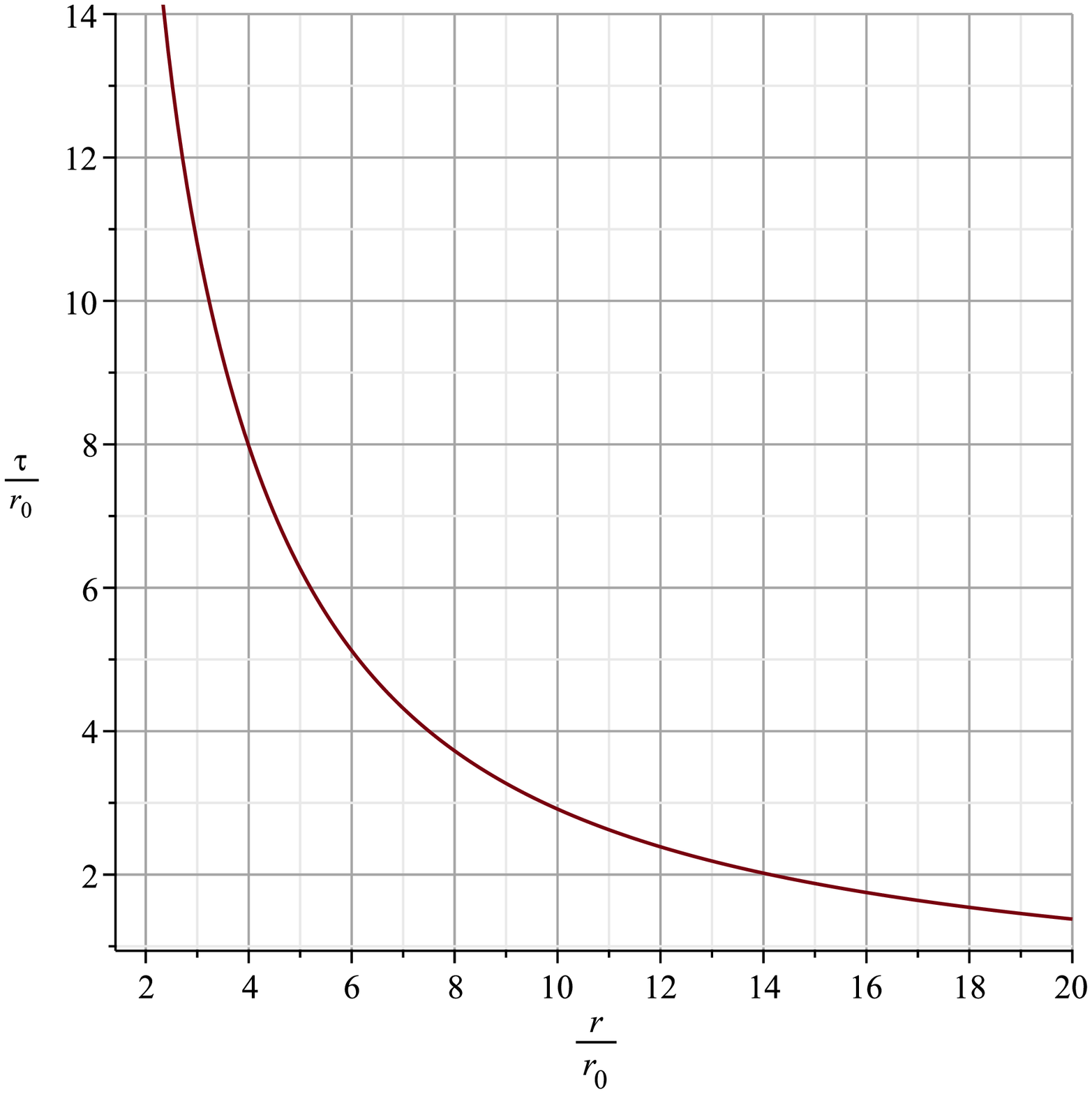}
 \label{3b}}
  \caption{\small{ In Fig (3a), the blue arrows represent the vector field $n$ on a portion of the $(x-\theta)$ plane for the BD-BHs in MG with $(q=r_{0}, c=1, c_{1}=-1/r_{0}, c_{2}=1, P=0.029/r_{0}^{2})$ where $(x=r/r_{0})$. The ZP is located at $(x,\theta)=(8.186, \pi/2)$ in the circular loop ($C_{1}$). The contours $C_1$ (purple loop) and $C_2$ (blue loop) are two closed loops, with $C_1$ enclosing the ZP while $C_2$ does not. The plot of the curve of equation (26) is shown in Fig (3b).}}
 \label{3}
 \end{center}
 \end{figure}
\newpage
\subsection{Bardeen AdS black holes in 4D Einstein-Gauss-Bonnet gravity}
A 4D Einstein-Gauss-Bonnet gravity BH is a type of BH that arises in a modified theory of gravity that includes a Gauss-Bonnet term in the action. This term is usually relevant only in higher dimensions, but some researchers have proposed a way to make it effective in four dimensions by re-scaling the coupling parameter. This theory is intended to bypass Lovelock's theorem, which states that the only theory of gravity with second-order field equations in four dimensions is general relativity \cite{30a}.
In this section, we discuss the GB-corrected BD-AdS BH solution in 4D space-time. The solution does not have any singularities. Overall, this study provides a new perspective on the topological GB model in 4-dimensional gravity \cite{31},
\begin{equation}\label{27}
f(r)=1+\frac{r^2}{2\alpha}\bigg(1\pm\sqrt{1+4\alpha\bigg[\frac{2M}{(r^2+q^2)^{3/2}}-\frac{1}{\ell^2}\bigg]}\bigg)
\end{equation}
When $\alpha\rightarrow0$, this solution becomes the BD-AdS BH, and the limit $q\rightarrow 0$ corresponds to the GB AdS-Schwarzschild solution. Solving the horizon condition $f(r_+)=0$  provides the mass in terms of its horizon radius. Therefore, the mass and entropy of this BH can be determined,
\begin{equation}\label{28}
M =\frac{4 \left(q^{2}+r^{2}\right)^{\frac{3}{2}} \pi  P \left(r^{4}+\frac{3 \left(r^{2}+\alpha \right)}{8 \pi  P}\right)}{3 r^{4}}
\end{equation}
and
\begin{equation}\label{29}
S =\pi  r^{2}+2 \pi  \alpha  \ln \! \left(r^{2}\right)
\end{equation}
Now, we obtain the generalized Helmholtz free energy for the BD-AdS BH in 4D EGB gravity as follows,
\begin{equation}\label{30}
\mathcal{F} =\frac{4 \left(q^{2}+r^{2}\right)^{\frac{3}{2}} \pi  P \left(r^{4}+\frac{3 \left(r^{2}+\alpha \right)}{8 \pi  P}\right)}{3 r^{4}}-\frac{\pi  r^{2}+2 \pi  \alpha  \ln \! \left(r^{2}\right)}{\tau}
\end{equation}
Here, we calculate $\tau$ with respect to equation (30) and we will have,
\begin{equation}\label{31}
\tau =\frac{4 \pi  r^{4} \left(r^{2}+2 \alpha \right)}{\sqrt{q^{2}+r^{2}}\, \left(8 P \,r^{6} \pi -2 r^{2} q^{2}+r^{4}-4 q^{2} \alpha -\alpha  r^{2}\right)}
\end{equation}

\begin{figure}[h!]
 \begin{center}
 \subfigure[]{
 \includegraphics[height=5.5cm,width=6cm]{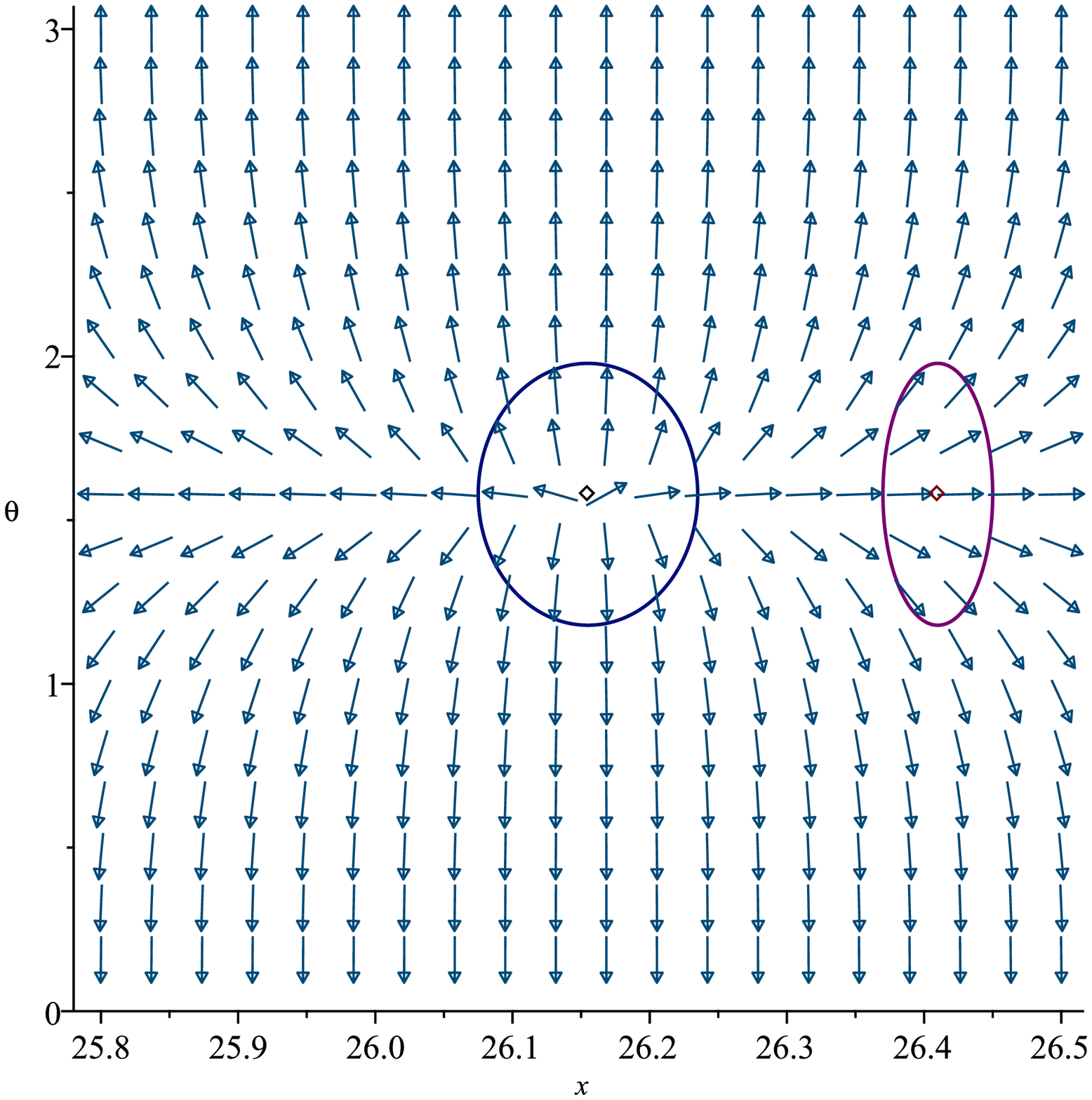}
 \label{4a}}
 \subfigure[]{
 \includegraphics[height=5.5cm,width=6cm]{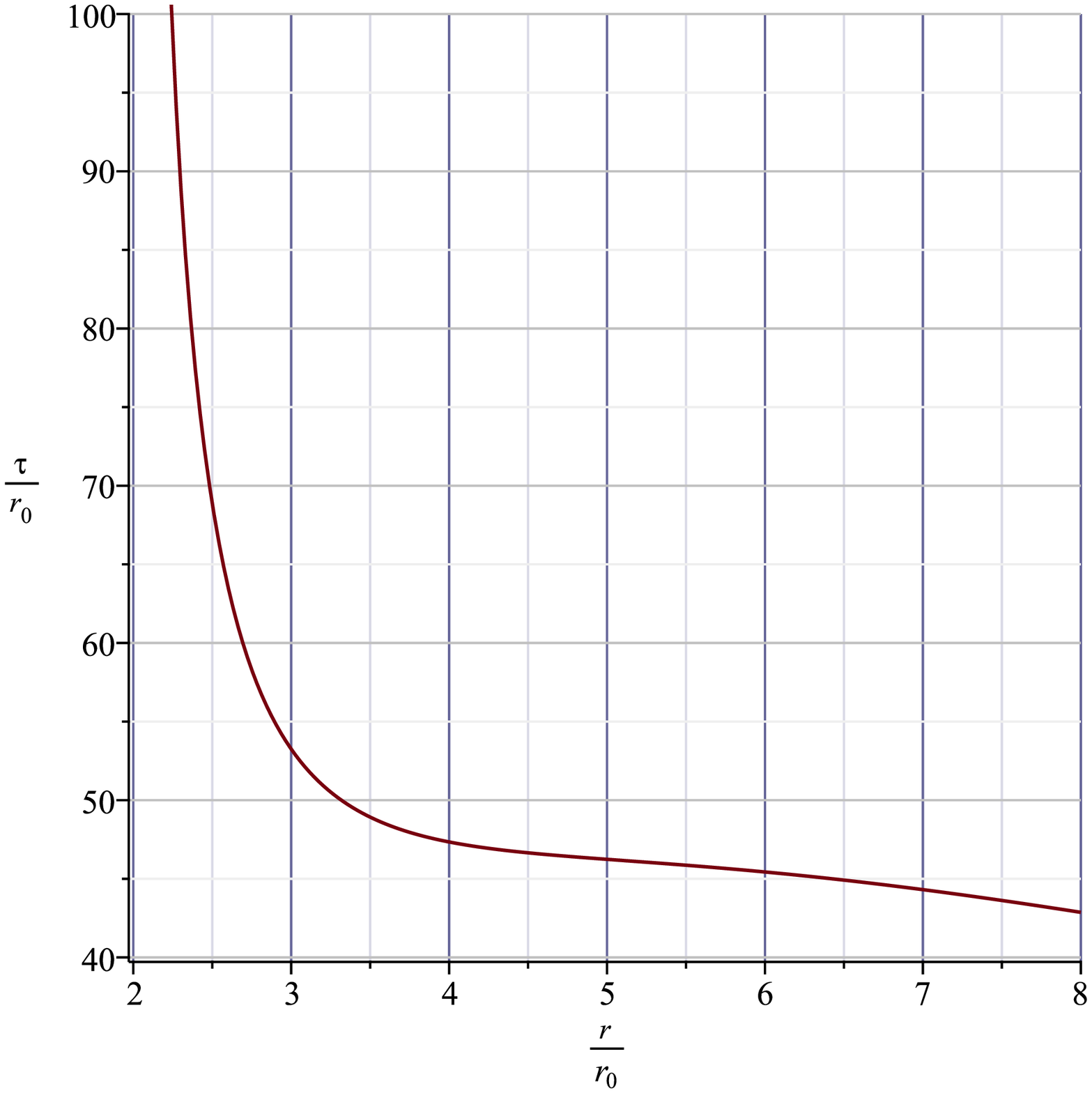}
 \label{4b}}
  \caption{\small{In Fig (4a), the blue arrows represent the vector field $n$ on a portion of the $(x-\theta)$ plane for the BD-AdS BH in 4D EGB Gravity with $(\alpha=r_{0}^{2}, g=r_{0}, P=0.0009/r_{0}^{2})$, where $(x=r/r_{0})$. The ZP is located at $(x,\theta)=(26.155,\pi/2)$ in the circular loop ($C_{1}$). The contours (blue loop) and $C_2$ (purple loop) are two closed loops, with $C_1$ enclosing the ZP while $C_2$ does not. The plot of the curve of equation (31) is shown in Fig (4b).}}
 \label{3}
 \end{center}
 \end{figure}
\section{Discussion}
We have analyzed Figure (1), which pertains to the structure of BD-AdS BHs. It is divided into two parts. The first part displays the normalized field lines. As depicted in the figure, there is only one zero point, which represents the only topological charge of this structure. It is proportional to the winding number and is situated inside the contour $(C_1)$ at the coordinates $(x, \theta) = (0.913, \pi/2)$.
We considered free parameters $(q=r_{0}, P=0.085/r_{0}^{2})$ to draw these contours. In Fig (1b), we plotted the curve related to equation (15) for different values of the free parameters.
As shown in Fig (1b), there is only one on-shell BH for arbitrary values of $(\tau)$ for BD-AdS BHs. Without loss of generality, we analyzed the topological properties of these BHs in this paper by choosing ($\tau=20r_{0}$) for regular BD-AdS BHs in the Kiselev’s model of quintessence and BD-AdS BHs in 4D EGB gravity, as well as ($\tau=30r_{0}$) for BD-BHs in MG. The results of this study on a BH indicate that the only characteristic of a positive topological charge of one is the zero point inside the contour ($C_1$), while the contour ($C_2$) that does not contain a zero point has zero topological charges. The content discusses the stability of the BH by examining the winding numbers and specific heat capacity. The positive value of the winding numbers suggests that the on-shell BH is thermodynamically stable, which can be confirmed by calculating the specific heat capacity. As there is only one on-shell BH, the topological number will be equal to the winding number, which is 1.
This means that there is only one stable on-shell BH, which is equivalent to a topological number equal to a positive winding number for all BH structures ($W=\omega=+1$). This indicates that the topological number may be unrelated to the components and parameters that describe different BD BH structures, such as Kiselev, MG, and EGB Gravity. We have plotted Figures (2, 3, 4) for regular BD-AdS BHs in Kiselev's model of quintessence, BD BHs in MG, and BD-AdS BHs in 4D EGB Gravity, respectively, following the same approach.
We plotted the normalized field vectors $n$ for these BHs in Fig. (2a, 3a, 4a). As shown in the figure, there is only one zero point located at $(x,\theta)=(1.175, \pi/2)$, $(x,\theta)=(8.186, \pi/2)$, and $(x,\theta)=(26.155,\pi/2)$ inside the contour $(C_1)$.
The contours $(C_1)$ and $(C_2)$ are also visible in Fig (2a, 3a, 4a), with the contour $(C_1)$ containing the only topological charge that is proportional to the winding numbers calculated for the mentioned BHs. We considered free parameters as $(q=r_{0}, c=1, c_{1}=-1/r_{0}, c_{2}=1, P=0.029/r_{0}^{2})$ for regular BD-AdS BHs in the Kiselev's model of quintessence, $(q=r_{0}, c=1, c_{1}=-1/r_{0}, c_{2}=1, P=0.029/r_{0}^{2})$ for BD BHs in MG, and $(\alpha=r_{0}^{2}, g=r_{0}, P=0.0009/r_{0}^{2})$ for the BD-AdS BHs in 4D EGB gravity to draw these contours.
In Fig (2b, 3b, 4b), we plotted the curve related to equations (21, 26, 31) for various values of the free parameters for the mentioned BHs. As $\tau$ decreases monotonically with the horizon radius $r$, it can be concluded that there is only one on-shell BH for an arbitrarily fixed $\tau$, and no phase transition occurs.
As shown in Fig (2b, 3b, 4b), there is only one on-shell BH for arbitrary values of ($\tau$).

\section{Conclusion}
We applied the generalized off-shell Helmholtz free energy method to investigate the thermodynamics of Bardeen black holes (BD BHs) from a topological viewpoint using Duan’s topological current $\phi$-mapping. We explored different types of BD BHs, such as regular BD-AdS BHs, BD-AdS BHs with Kiselev’s quintessence model, BD-BHs in massive gravity (MG), and BD BHs in 4D Einstein-Gauss-Bonnet (EGB) gravity. We showed that these BHs belong to one topological class (TC), i.e., TC is +1 for all cases, and that adding MG or GB terms, etc., does not affect the topological numbers. We summarized the results in Table 1. These results motivate future investigations into the TCs of BHs in rainbow gravity, as well as in stringy and supergravity BHs.

\begin{center}
\begin{table}
  \centering
\begin{tabular}{|p{8cm}|p{8cm}|}
  \hline
  \centering{Case}  & \hspace{2cm} Topological Numbers \\[3mm]
   \hline
  \centering{BD-AdS black holes} & \hspace{3cm} $W=+1$ \\[3mm]
   \hline
  \centering{BD-AdS black holes in Kiselev’s model of quintessence} & \hspace{3cm} $W=+1$ \\[3mm]
   \hline
   \centering{BD-AdS black holes in massive gravity}  & \hspace{3cm} $W=+1$ \\[3mm]
    \hline
   \centering{BD-AdS black holes in 4D Einstein-Gauss-Bonnet gravity} & \hspace{3cm} $W=+1$\\[3mm]
  \hline
\end{tabular}
\caption{Summary of the results.}\label{20}
\end{table}
 \end{center}

\end{document}